\documentclass[10pt,draft]{dis03}
\usepackage{epsf,amsmath}

\begin{document}

\title{DIS and ``elastic'' diffraction 
\thanks{Supported by Deutsche Forschungemeinschaft, Schi 189/6-2}
\thanks{Presented at DIS 2003, St. Petersburg, Russia, April 2003}}

\author{Dieter Schildknecht \\
Faculty for Physics, University of Bielefeld \\
D-33501 Bielefeld, Germany\\
E-mail: Dieter.Schildknecht@physik.uni-bielefeld.de }

\maketitle

\begin{abstract}
\noindent 
\!The QCD-based generalized vector dominance-color-dipole picture \!(GVD-CDP) 
provides a coherent picture of low-$x$ DIS, deeply virtual Compton
scattering and light as well as heavy vector-meson production.
\end{abstract}

\section*{} 
More than thirty years ago, it was suggested 
\cite{1,2}
that deeply inelastic scattering of electrons from nucleons at low 
$x \cong Q^2/W^2$ was to be understood in terms of the forward-scattering 
amplitude of a continuum of massive hadronic vector states the 
photon was supposed to virtually dissociate into, or, in modern jargon, 
``to fluctuate into''. A mass dispersion relation was set up to 
quantitatively formulate this picture of generalized vector dominance (GVD). 
Concerning the energy dependence of the virtual photoabsorption cross 
section, $\sigma_{\gamma^* p} (W^2, Q^2)$, not surprisingly, the simplest 
assumption was adopted, an energy dependence determined by what nowadays is 
called the ``soft'' Pomeron. 

When HERA came into operation about ten years ago, an appreciable fraction of the
hadronic events (``large-rapidity gap events'') was found to 
show typical features of 
diffractive production, whereby including the production of a massive continuum
of hadronic states as well as the elastic production of the vector mesons, 
$\rho^0, \omega, \phi, J /\psi$ and $\Upsilon$. The energy dependence found in
diffractive production and for the total photoabsorption cross section, however, 
for sufficiently large $Q^2$, turned out to be much stronger than expectations
based on the soft Pomeron relevant for (real) photoproduction and 
hadron-hadron-scattering. 

The experimental data on $\sigma_{\gamma^* p} (W^2, Q^2)$ for 
$x \ll 1$, in good approximation, in a model-independent analysis were found 
\cite{3}
to lie on a single curve when plotted against the variable \cite{3}
\begin{equation}
\eta = \frac{Q^2 + m^2_0}{\Lambda^2 (W^2)}, 
\label{(1)}
\end{equation}
where \cite{4}
\begin{eqnarray}
m^2_0 & = & 0.15 \pm 0.04 \, {\rm GeV}^2   \nonumber \\
\Lambda^2 (W^2) & = & B \left( \frac{W^2}{W^2_0} + 1 \right)^{C_2} 
\label{(2)}
\end{eqnarray}
with 
\begin{eqnarray}
B & = & 2.24 \pm 0.43 \, {\rm GeV}^2 , \nonumber \\
W^2_0 & = & 1081 \pm 124 \, {\rm GeV}^2 , \label{(3)} \\
C_2 & = & 0.27 \pm 0.01 . \nonumber 
\end{eqnarray}

The modern QCD-based analysis of low-$x$ DIS describes the Pomeron as 
a two-gluon-exchange object \cite{5}. The virtual photon at low $x$ 
fluctuates into a $(q \bar q)^{J=1}$ (vector) state, as in GVD, and the 
$(q \bar q)^{J=1}$ color dipole, in the virtual Compton-forward-scattering 
amplitude, interacts with the proton via two-gluon exchange. A detailed analysis
reveals that the resulting forward-Compton amplitude embodies a structure of the 
form the aforementioned mass-dispersion relation \cite{6,7}. The quark
propagators in the quark loop of the two-gluon exchange become transmogrified
into propagators of $(q \bar q)^{J=1}$ vector states of mass 
\begin{equation}
M^2_{q \bar q} = \frac{\vec k^2_\bot + m^2_q}{z(1-z)} , 
\label{(4)}
\end{equation}
where $k_\bot$ and $m_q$ refer to transverse three-momentum and mass of the 
quark (antiquark) and $z$ to the fraction of the light-cone momentum of the 
photon carried by the quark. Our formulation \cite{3} for the 
total photoabsorption cross section in terms of the photon-lightcone wave 
function and the color-dipole cross section explicitly incorporates \cite{8}
the spin $J=1$ nature of the color dipole in the color-dipole 
forward-scattering amplitude, i.e. the dipole-cross section refers to the 
scattering of $(q \bar q)^{J=1}$ (vector) states 
(generalized vector dominance - color dipole picture, GVD-CDP).
This allowed us to derive sum rules \cite{8} that express the longitudinal 
and transverse total photoabsorption cross section as an appropriate integral 
over the mass of the diffractively produced $(q \bar q)^{J=1}$ continuum, 
including the low-lying discrete vector-meson states. A direct test of the 
sum rules requires the extraction of the $(q \bar q)^{J=1}$ component in the 
experimental data for diffractive production that has not been accomplished 
so far. The underlying structure of the theory may be tested, however, by 
confronting the parameter-free predictions for vector-meson production with 
the experimental data, and we will come back to that below. 

In the QCD-based GVD-CDP, the total photoabsorption cross section in the 
limits of $\eta \ll 1$ and $\eta \gg 1$, is given by \cite{3}
\begin{equation}
\sigma_{\gamma^* p} (W^2 , Q^2) = \frac{\alpha R_{e^+ e^-}}{3\pi} 
\sigma^{(\infty)} \left\{ \begin{matrix}\ln (1/\eta), & &
(Q^2 \ll \Lambda^2 (W^2)), \\
1 / 2 \eta, & & (Q^2 \gg \Lambda^2 (W^2)),\end{matrix} \right.
\label{(5)}
\end{equation}
and $\Lambda^2 (W^2)$ becomes identified with the average or effective value of 
the gluon transverse (three) momentum absorbed by the quark (antiquark), via 
\cite{4}
\begin{equation}
\langle \vec l^2_\bot \rangle = \frac{1}{6} \Lambda^2 (W^2) . 
\label{(6)}
\end{equation}

An important comment concerns to gluon structure function. For 
$Q^2 \gg \Lambda (W^2)$ in a dual description, one may alternatively describe 
$\sigma_{\gamma^* p} (W^2 , Q^2)$ in terms of $\gamma^*$-gluon scattering, and 
introduce the concept of a gluon density or gluon structure function. 
Accordingly, we have the identification \cite{9,10}
\begin{equation}
\frac{1}{8\pi^2} \sigma^{(\infty)} \Lambda^2 (W^2) \equiv \alpha_s (Q^2) x
g (x , Q^2) , 
\label{(7)}
\end{equation}
where $W^2 = Q^2 / x$. It is frequently argued that the gluon-structure 
function should eventually stop its growth with decreasing $x$ at fixed 
$Q^2$, because of the unitarity restrictions on the total photoabsorption 
cross section. From our point of view, there is a peaceful coexistence with 
unitarity of an ever rising gluon density. Indeed, at any fixed $Q^2$, for 
sufficiently large energy (i.e. sufficiently small $x$) the 
interpretation of $\sigma^{(\infty)} \Lambda^2 (W^2)$ as a gluon density
simply breaks down, and the logarithmic domain in (\ref{(5)}) takes over, and
finally, at any fixed $Q^2$, we reach the (non-perturbative) saturation 
limit of photoproduction \cite{4}
\begin{equation}
\lim_{{W^2\rightarrow\infty}\atop{Q^2 \, {\rm fixed}}}
\frac{\sigma_{\gamma^* p} (W^2 , Q^2)}{\sigma_{\gamma p} (W^2)} = 1. 
\label{(8)}
\end{equation} 

As mentioned, a direct test of the QCD-based GVD-CDP is provided by comparing its
parameter-free predictions for $(q \bar q)^{J=1}$ vector-state forward production
with experiment. So far, data are only available for the production of 
vector mesons, since an extraction of the $J=1$ part in the diffractive 
continuum has not been carried out. As an example, in fig. 1, we show the 
GVD-CDP predictions \cite{11} compared with the data for $\rho^0$ production, 
whereby a constant slope of $b = 7.5 \, {\rm GeV}^{-2}$ was inserted. The 
prediction in fig. 1 is obtained by applying quark-hadron duality to 
the diffractively produced continuum in the $\rho^0$-mass region. While for 
$\rho^0$ production the approximation by massless quarks is relevant, i.e. 
$M^2_{\rho^0} \gg 4 m^2_q \cong 0$, in the case of $J / \psi$ and $\Upsilon$, the 
approximation $M^2_V \cong 4 m^2_q$ must be used. From the light-cone wave 
functions, at threshold, for $z (1-z) = \frac{1}{4}$, one finds the 
substitution prescription \cite{11}
\begin{equation}
Q^2 \rightarrow Q^2 + 4 m^2_q 
\label{(9)}
\end{equation}
to pass from the massless-quark case to the massive-quark case. Again 
applying quark-hadron duality, one finds an enhancement, \cite{11}
\begin{equation}
E^{(V)} \equiv \frac{R^{(\rho^0)}}{R^{(V)}} \frac{\sigma_{\gamma^* p 
\rightarrow Vp}}{\sigma_{\gamma^* p \rightarrow \rho^0 p}}, (R^{(\rho^0)}/
R^{(J/\psi)} = \frac{9}{8}, R^{(\rho^0)}/R^{(\Upsilon)} = \frac{9}{2}),
\label{(10)}
\end{equation}
of $E^{(J/\psi)} \cong 1.4$ of $J/\psi$ photoproduction relative to 
$\rho^0$ production at $Q^2 \simeq M^2_{J/\psi}$, and an 
enhancement of $E^{(\Upsilon)} \cong 2.7$ for $\Upsilon$ photoproduction
relative to $\rho^0$ production at $Q^2 \cong M^2_\Upsilon$, consistent with 
experimental results \cite{12}.
Concerning the energy dependence, in fig.2, we show the GVD-CDP prediction 
\cite{11} for $\delta^{(V)}$, where our results are adapted to the 
experimentlists' fits of the form $W^{\delta (V)}$. Finally, in fig. 3, we show 
our parameter-free predictions \cite{11} for the $Q^2$ dependence of DVCS
compared with HERA data \cite{13}. 

After many years of experimental and theoretical efforts, a coherent picture
of DIS in the low-$x$ region, DVCS and vector-meson production has emerged for 
all $Q^2 \ge 0$. The $Q^2$ dependence and the relative normalization of the 
cross sections is well understood from the QCD-based 
GVD-CDP. The scale of the transition from a hard to a soft energy 
dependence is provided by $\Lambda^2 (W^2)$, where $\Lambda^2 (W^2)/6$ is
identified with the effective gluon transverse momentum. The ab initio
prediction of the power of $W^2$ in $\Lambda^2 (W^2)$ is beyond the scope of 
our present understanding.

\section*{Acknowledgements} 
It is a pleasure to thank G. Cvetic, M. Kuroda, B. Surrow and
M. Tentyukov for a fruitful collaboration that lead to the result 
presented here.

\vspace{2cm}

\begin{figure}[htbp]\centering
\epsfysize=7.5cm
\centerline{\epsffile{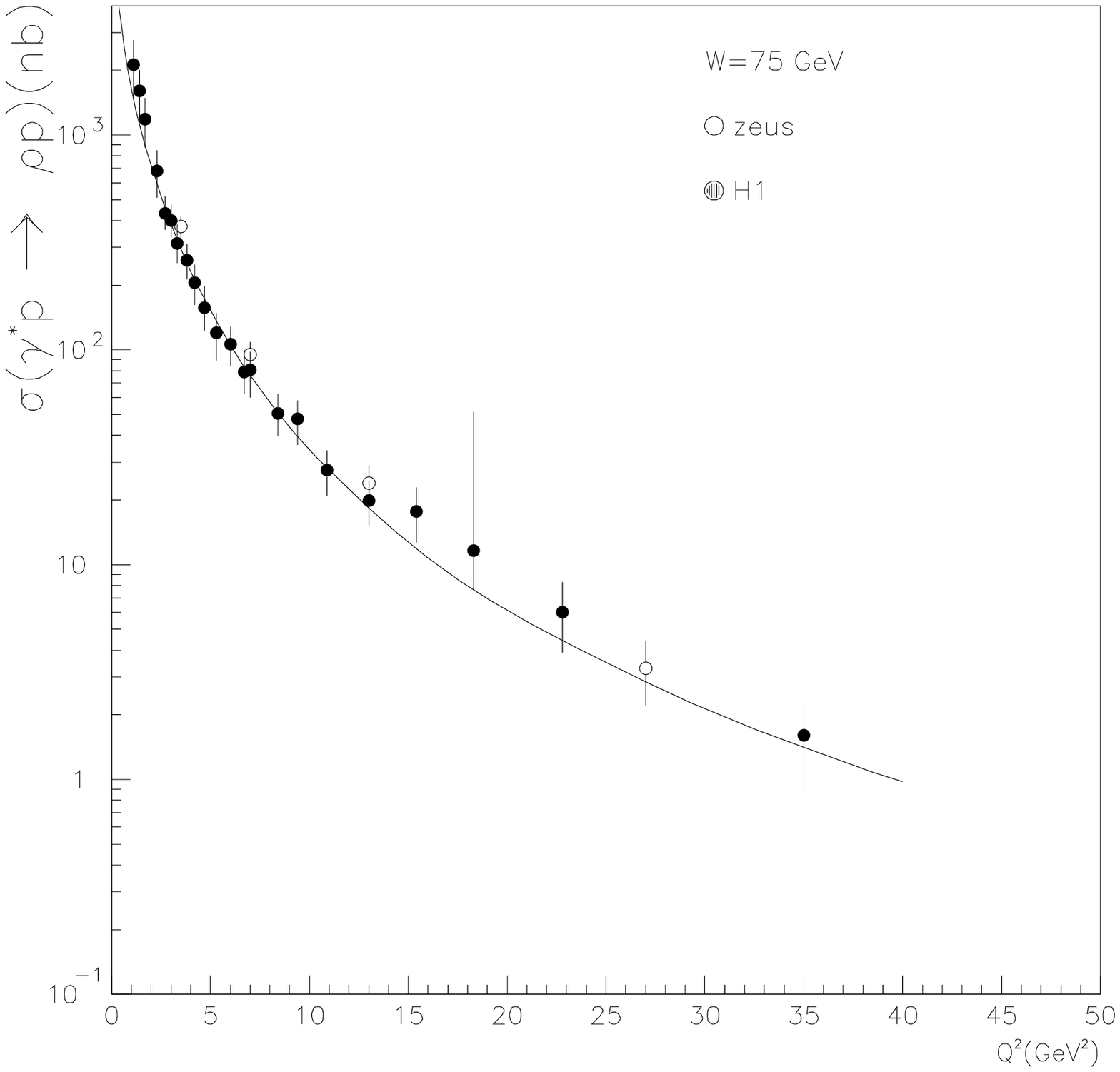}}
\caption[*]{
The $Q^2$ dependence of $\rho^0$ production, $\gamma^*p \to
\rho^0 p$, at fixed $W = 75~GeV$, compared with the predictions from
the QCD-based GVD-CDP.
}
\end{figure}

\begin{figure}[t]\centering
\epsfysize=6.5cm
\centerline{\epsffile{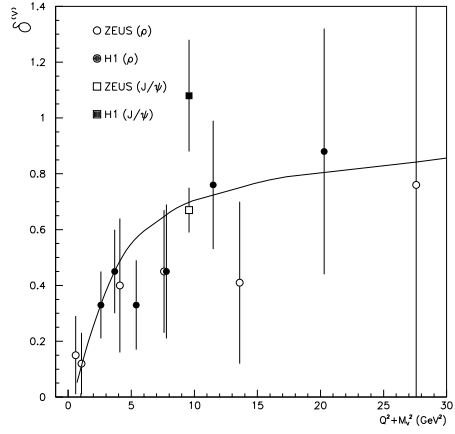}}
\caption[*]{ 
The exponent $\delta^{(\rho^0)}$ in a parameterization of the 
energy-dependence of the experimental cross section by 
$W^{\delta^{(V)} (Q^2 + M^2_V)}$ compared
with the predictions from the QCD-based GVD-CDP.
}
\end{figure}
\vspace*{-1cm}

\begin{figure}[b]\centering
\epsfysize=6.5cm
\centerline{\epsffile{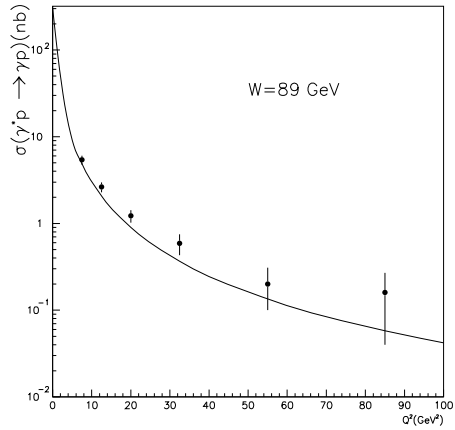}}
\caption[*]{
The $Q^2$ dependence of DVCS at the energy of $W=89$ GeV compared with the 
predictions from the QCD-based GVD-CDP.
}
\end{figure}

\vfill\eject

\end{document}